\documentclass[
    ,final            
  ]
  {aipproc}

\layoutstyle{8x11double}

\usepackage{xspace}

\newcommand{\dzero}     {D0}
\newcommand{\met}       {\mbox{$\not\!\!E_T$}\xspace}

\begin{document}

\title{Search for charged Higgs Bosons at \dzero}


\author{Yvonne Peters for the D\O\ Collaboration}{
  address={ University of Wuppertal \\
Gaussstrasse 20, 42097 Wuppertal \\
E-mail: peters@fnal.gov} }

\begin{abstract}
In both Supersymmetry and in generic Two Higgs Doublet models (2HDM), the
charged Higgs boson $H^{\pm}$  exhibits a unique phenomenological
signature. We report on a  search for charged Higgs bosons, performed
using $0.9{\rm ~fb}^{-1}$  of data collected with the \dzero~detector at the
Fermilab Tevatron {\mbox{$p\bar p$}}\ collider with a center-of-mass energy of {\mbox{$\sqrt{s}$
=\ 1.96\ TeV}}. No evidence for a charged Higgs
boson is found and we set limits on its production cross section or the branching fraction. 
\end{abstract}

\maketitle

\section{Introduction}
In the Standard Model (SM) of particle physics one SU(2) doublet is needed to 
induce electroweak symmetry breaking, leading to one observable particle
called the Higgs boson. So far no hint for the Higgs boson has been observed
experimentally. Furthermore, there is no reason to believe that the electroweak
symmetry breaking is caused by only one Higgs field.  
A second  SU(2) doublet is introduced in two-Higgs-doublet models (2HDM).
This leads to five physical Higgs bosons, three of which are neutral ($h^0$,
$A^0$ and $H^0$) and the
other two carry electric charge ($H^\pm$).  

In the 2HDM models, the coupling of fermions to the Higgs fields is not
specified by the theory. The only
requirement is to avoid flavor-changing neutral currents (FCNC). Three types
of 2HDM are considered~\cite{higgs_hunter, higgs_type3}. In  Type~I 2HDM, only one Higgs doublet couples to
fermions, in Type~II one doublet couples to up-type, the other to down-type
fermions. In Type~III models, both doublets couple to fermions and additional
methods are introduced to suppress FCNCs, for example, by exploiting the small mass of
first and second generation quarks.  A very prominent candidate for physics
beyond the SM is Supersymmetry (SuSy). In SuSy, and in the simplest sypersymmetric extension
of the SM, the Minimal
Supersymmetric Standard Model (MSSM), only Type~II 2HDM models can be
realized. At tree-level MSSM two parameters are
relevant: the mass of the $H^{\pm}$ and $\tan \beta$, the ratio of the vacuum expectation values of
the two Higgs fields. 

We present two searches for charged Higgs bosons in the top quark sector. The top quark,  the fermion with the highest mass, has
the strongest coupling to the Higgs sector of all known particles. In the first
analysis we search for  a charged Higgs boson heavier  than the top quark in the single top production
channel~\cite{stopPRL}. The second search explores the production of 
 $H^{\pm}$ 
lighter  than the top quark in the top quark pair ($t\bar{t}$) production
channel.

\section{A search for heavy charged Higgs bosons}
In 2006, the first evidence for single top quark
production was reported by \dzero~\cite{stopPRL}. While for this measurement
the $s$- and $t$-channels were considered together, it is worth to study both
production mechanisms separately~\cite{stopPRD}, especially since the  $s$-
and $t$-channel show different sensitivity to new physics. For example, 
searches for $W'$ bosons~\cite{Wprime}, anomaleous $Wtb$ couplings~\cite{Wtb} or the
search  for charged Higgs bosons~\cite{Hplus_heavy} are performed in the $s$-channel,
the latter search being
presented here. 

\begin{figure}[h]
  \includegraphics[height=.07\textheight]{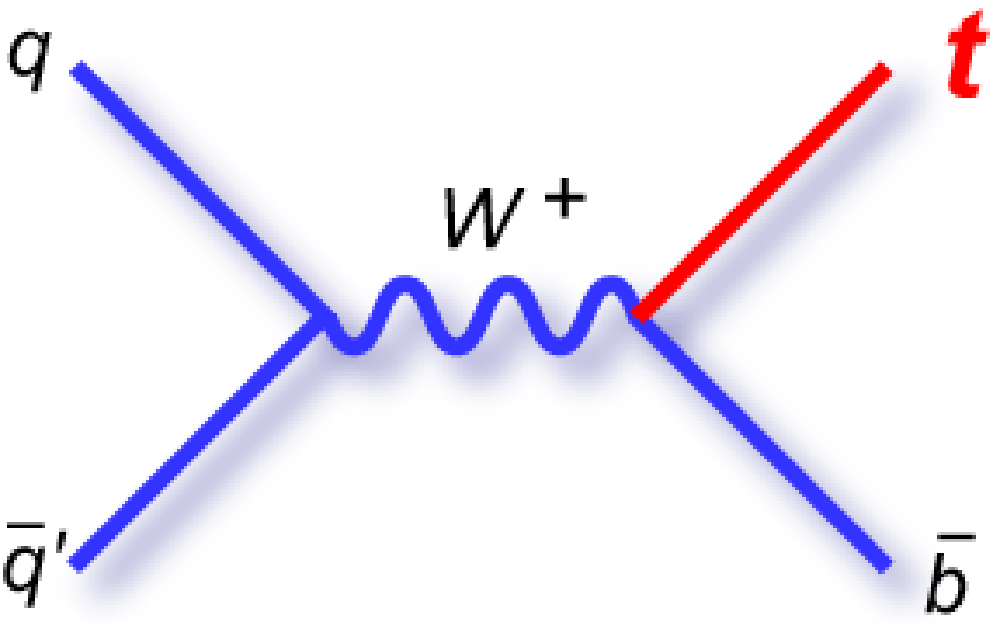} \hspace{0.3cm}
  \includegraphics[height=.07\textheight]{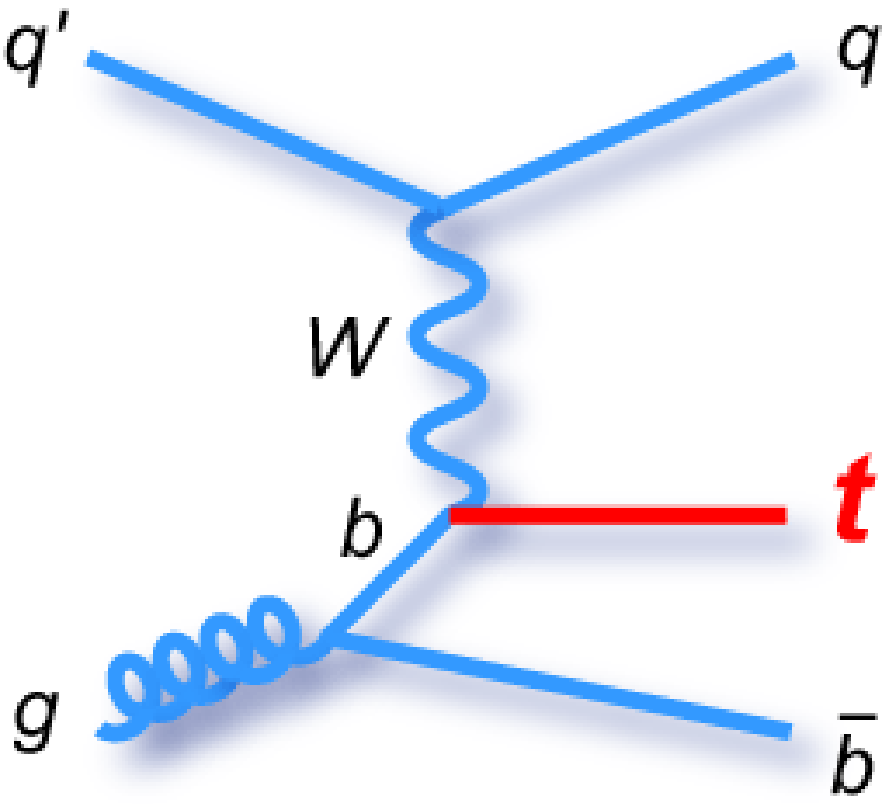} 
  \caption{$s$- and $t$-channel diagram of single top quark production.\label{single_top}}
\vspace{-1cm}
\end{figure}

The branching fraction ${\mathcal B}(H^{+} \rightarrow t\bar{b})$ is close to unity for a large range of $\tan \beta$. We study the cross section times branching fraction $\sigma(q\bar{q}'
\rightarrow H^{+})\times {\mathcal B}(H^{+} \rightarrow t\bar{b})$ (referred
to as $\sigma
\times {\mathcal B}$)
in the $s$-channel (Fig.~\ref{single_top} left). The $t$-channel
(Fig.~\ref{single_top} right) is not relevant  due to the small
couplings of $H^{+}$ to light quarks. 

In the $s$-channel, the decays \mbox{$H^{+} \rightarrow t\bar{b}$} and \mbox{$W^{+}
\rightarrow t\bar{b}$} result in the same final state. Therefore the same event
selection as for the single
top production cross section measurement is used for the $H^{+}$  search. While for the latter subsamples with
jet multiplicities of $2$, $3$ and $4$ jets are considered, only events with
exactly two jets are considered for the $H^{+}$ search. This is the
subsample with the highest sensitivity for $s$-channel single top quark production. Events where exactly one isolated electron (muon) with $E_T>15$~GeV ($E_T>18$~GeV)
and pseudorapidity $|\eta|<1.1$ ($|\eta|<2.0$) and transverse missing energy \met
of $15{\rm~GeV}<\met<200{\rm~GeV}$ are selected. Additionally, one jet
has to fulfil $p_T>25$~GeV and $|\eta|<2.5$, the other $p_T>20$~GeV and
$|\eta|<3.4$ requirements. At least one of the jets must be identified as a $b$-jet. 

The $b$-jet identification is done using  a
neural-network tagging algorithm~\cite{btagging}. It combines
variables characterizing the presence and properties of tracks with high
impact parameter and seconday vertices inside a jet. In the simulation,
the events are weighted by a probability for each jet to be $b$-tagged, derived
from data control samples. 

The simulation of the charged Higgs signal is performed by setting the coupligs
such that pure chiral samples are produced. The combination  of different
proportions of purely left and right-handed samples is used to simulate the
three different 2HDM types, as the chriality and up/down-type fermions are correlated. 

In order to discriminate charged Higgs production from SM processes, the
region of  charged Higgs masses,
$m_{H^{+}}>m_t$, is explored. We build the reconstructed invariant mass of the two jets
and the $W$-boson $M({\rm jet1, jet2,} W)$ for all Monte Carlo and
data samples. A binned likelihood in the invariant mass is constructed. 

No excess of data over the SM prediction can be observed. Therefore, we set
upper limits on $\sigma
\times {\mathcal B}$ in all
three 2HDM models using Bayesian statistics with a flat  prior  in the signal cross section. 
The limits can be parametrized as a function of $m_{H^{+}}$ and $\tan \beta$ in
case of the Type~I and Type~II 2HDM, and as function of $m_{H^{+}}$ and the
top-charm quark mixing parameter $\zeta$ in the Type~III 2HDM. The expected
and observed limits 
come out close to each
other. 

\begin{figure}
  \includegraphics[height=.19\textheight]{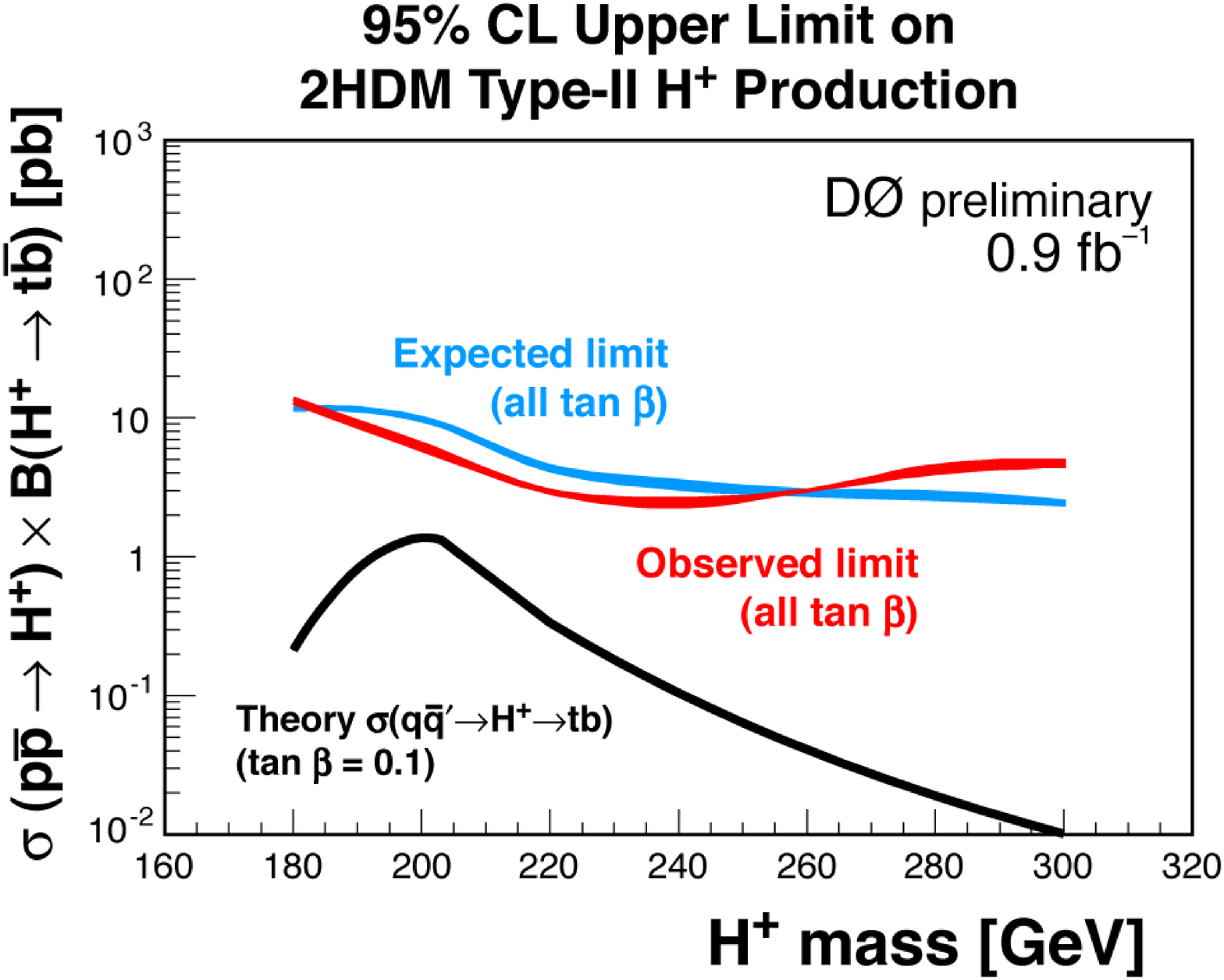} \hspace{0.1cm}
  \includegraphics[height=.19\textheight]{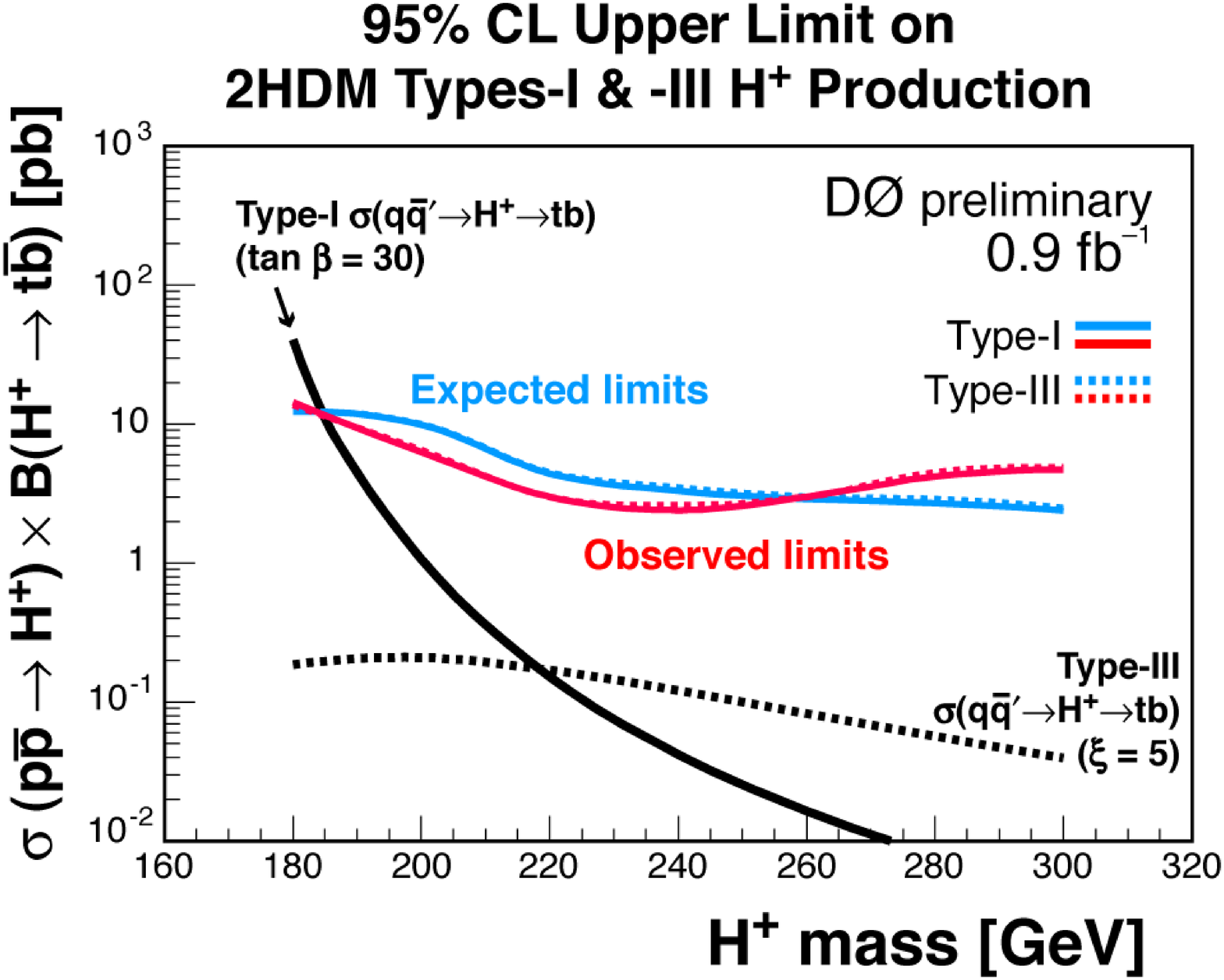} \hspace{0.1cm}
  \includegraphics[height=.17\textheight]{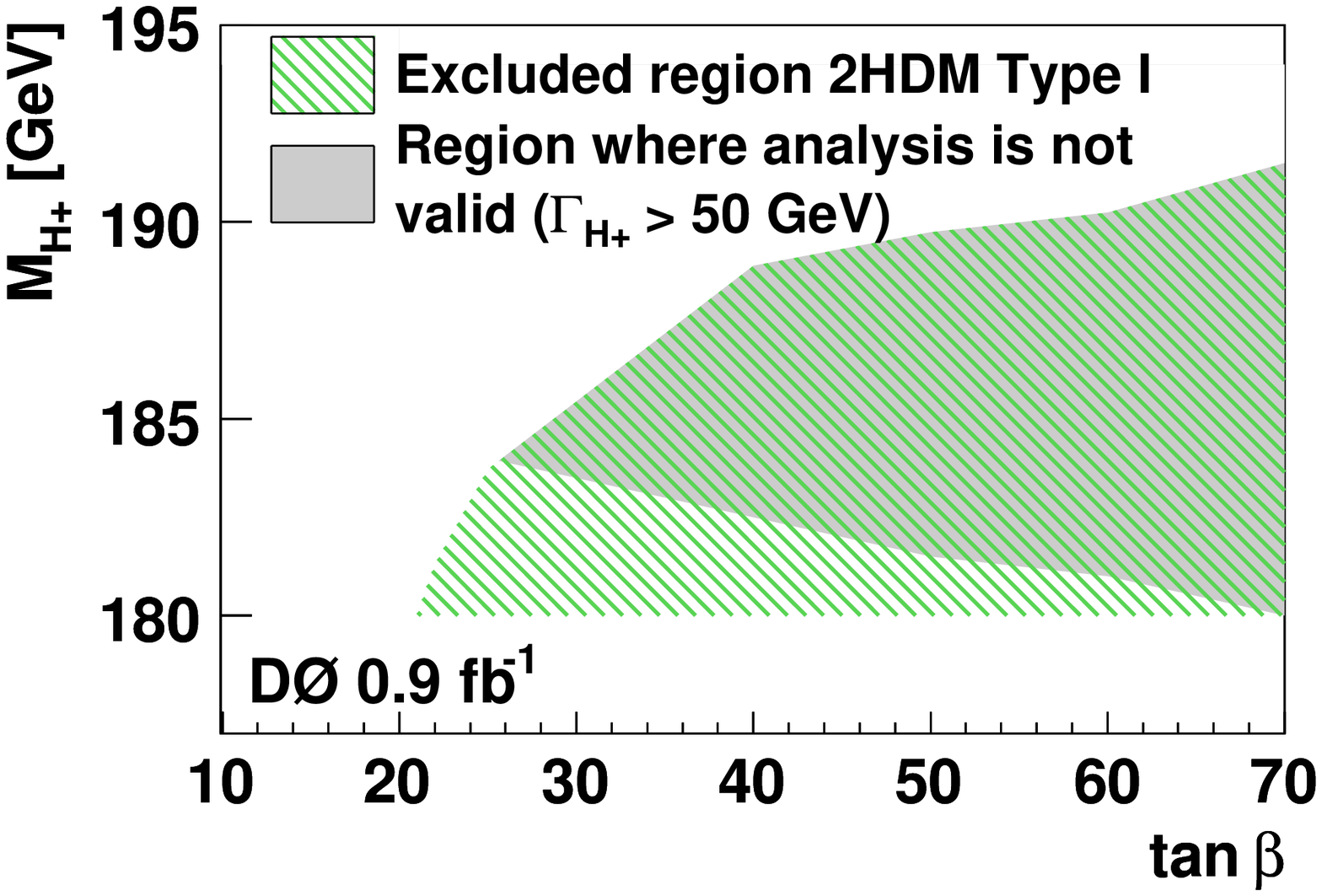}
  \caption{Expected and observed limtis for Type~II (left) and Type~I and~III (middle)
  2HDM models. The right plot shows the $95$~\% Cl. L. exclusion region in the
  ($\tan \beta$, $m_H^{+}$) space for the Type~I 2HDM model.
  \label{single_top_exclude}}
\end{figure}

The limits are compared to the expected signal cross section in the different
2HDM models. Figure~\ref{single_top_exclude} shows the expected and observed
limits together with the theoretical predictions for the Type~II (left) and Type~I
and Type~III models (middle). 
In Type~I, charged Higgs masses between 180 and 184~GeV can be
excluded in the \mbox{($\tan \beta$, $m_H^{+}$)} parameter space, as shown in
Fig.~\ref{single_top_exclude}~(right). In the Type~II and Type~III models, the analysis sensitivity is
not sufficient yet to exclude regions in the \mbox{($m_{H^{+}}$, $\tan\beta$)} or 
\mbox{($m_{H^{+}}$, $\zeta$)} parameter space, respectively. 

\section{A search for light charged Higgs bosons}
The $t\bar{t}$ production cross section at next-to-leading order is $6.8\pm0.6$~pb
at a top quark mass of $175$~GeV at the Tevatron~\cite{SMtheory}. This yields several thousands of
top quark pairs with the current integrated luminosity, large
$t\bar{t}$ yields after selection cuts  enabling precision measurements and
searches for new physics in the top quark sector. 

In the SM, the top quark decays into a $W$ boson and a $b$-quark. Different final
states are classified according to
the decay of the two $W$ bosons in the $t\bar{t}$ system. The final state in which both $W$'s decay into a lepton
(electron
or muon) is called a dilepton channel. If one $W$ boson decays into electron or
muon and one into quarks, the final state is called lepton plus jets
($\ell$+jets).

Requiring at least four jets with at least one of them
$b$-tagged in the $\ell$+jets channel, the  $t\bar{t}$ cross section is
measured with about $1~{\rm fb}^{-1}$ of data at a top quark mass of $175$~GeV as~\cite{hplus} 
\begin{eqnarray*}
\sigma(t\bar{t})_{\ell{\rm+jets}}\! =\!
  8.27^{+0.96}_{-0.95}{\rm~(stat+syst)}\pm 0.51{\rm~(lumi)~pb\,. \ \ \   } \nonumber
\end{eqnarray*}
In the dilepton final state we obtain~\cite{hplus} 
\begin{eqnarray*}
\sigma(t\bar{t})_{\rm dilepton}\! =\! 6.8^{+1.2}_{-1.1}{\rm~(stat)}^{+0.9}_{-0.8}{\rm~(syst)}\pm\! 0.4{\rm~(lumi)~pb}\,.\nonumber
\end{eqnarray*}
The requirement of at least four jets in the $\ell$+jets channel is chosen to
minimize the overlap between dileptonic and semileptonic final states. 

In the presence of
a charged Higgs boson lighter than the top quark, the decays $t\rightarrow Wb$
and $t\rightarrow H^{+}b$ can compete. Depending on  the  $H^{+}$
decay mode, for example a pure decay into $\tau \nu$ or $c\bar{s}$, the expected
yields in the various final states of a $t\bar{t}$ pair can significantly deviate from the SM
expectation.

Since a new decay mode of the top quark can change the measured cross section,
the ratio of the cross sections measured in the $\ell$+jets and dilepton channels 
\begin{eqnarray*} 
R_{\sigma}  =  \frac{\sigma( t\bar{t})_{\ell{\rm+jets}}}{\sigma(t\bar{t})_{\rm dilepton}} 
\end{eqnarray*}
 can be used to
  explore alternative models beyond the SM~\cite{hplus}. 
  By
  generation of pseudo-experiments taking into account the systematic
  uncertainties and their correlations we derive 
  $R_{\sigma}=1.21^{+0.27}_{-0.26}{\rm~(stat+syst)}$.

Any deviation from the SM value of 
  $R_{\sigma}=1$ can be a hint for new physics. For example, the branching
  ratio ${\mathcal B}(t\rightarrow bX)$ with $X$ being any  particle but the $W$
  boson can lead to $R_{\sigma}$ different from one.  In case of $X:=H^{+}$, the ratio of cross
  sections can be used to search for light charged Higgs bosons.

We study a simple model in which ${\mathcal B}(H^{\pm}\rightarrow cs)$ is
100~\%, the
charged Higgs boson mass is close to the $W$ mass, and therefore a 
 similar event kinematics of the  $t\rightarrow   bW$ and
$t\rightarrow   bH^{+}$ decays can be assumed. This leptophobic charged Higgs model is
interesting at low $\tan\beta$ in the MSSM for scenarios with large
suppression of the tauonic decay, for example,  from SuSy-breaking effects~\cite{carena}. Furthermore, in general
multi-Higgs-doublet models (MHDM), a leptophobic decaying charged Higgs is
possible for the full $\tan\beta$ range~\cite{Grossman}. For a $H^{\pm}$ mass of
$80$~GeV and $\tan \beta<3.5$, such a leptophobic charged Higgs boson could lead to
noticeable effects at the Tevatron~\cite{Akeroyd}. 

As $R_{\sigma}$ shows no deviation from one within its uncertainty, we set
upper limits on ${\mathcal B}(t \rightarrow H^{+}b)$ for a leptophobic
decaying charged Higgs boson with a mass of $80$~GeV. We use a frequentist
method from Feldman and Cousins~\cite{fc_limit}, resulting in an upper
observed limit of ${\mathcal B}(t \rightarrow H^{+}b)<0.35$ at $95$~\% C.L.,
and an expected limit of ${\mathcal B}(t \rightarrow H^{+}b)<0.25$ at $95$~\% C.L.

The earlier measurement of the ratio $1/R_{\sigma}$ from CDF with
$200{\rm~pb}^{-1}$ yields $1/R_{\sigma}=1.45^{+0.83}_{-0.55}$ and a limit on
${\mathcal B}(t\rightarrow Xb)$ of less than $0.46$ at $95$~\% C.L.~\cite{CDF_ratio} under the assumption of
the same acceptance for  \mbox{$t\rightarrow   bW$} and
\mbox{$t\rightarrow   bH^{+}$} decays. 

A dedicated search for light charged Higgs bosons between $80$ and $155$~GeV
 has been released by \dzero~recently~\cite{hplus_new}, where the yields in
 the dilepton, $\ell$+jets and $\tau$+lepton channels are used to search for the
 charged Higgs. 
 Limits on a
purely tauonic and a leptophobic charged Higgs model have been set, resulting
in \mbox{${\mathcal B}(t\rightarrow H^{+}b)<0.2$} in the tauonic model and  \mbox{${\mathcal B}(t\rightarrow
H^{+}b)<0.26$} in the leptophobic model for charged Higgs masses between $80$
 and $15$~GeV. 
The procedure is complementary to the consideration of
the cross section ratio.

\section{Conclusion and outlook}
We presented  searches for charged Higgs bosons in single top and top pair events at \dzero, with about
$1{\rm~fb}^{-1}$ of integrated luminosity. In the single top channel we
searched for heavy charged Higgs bosons, interpreted in terms of upper limits
on $\sigma \times {\mathcal B}$ for Type~I,~II and~III 2HDM models. First
exclusion limits have been set in the ($m_{H^{\pm}}$, $\tan\beta$) space for
Type~I models. 

 In the
$t\bar{t}$ production mode, we derived the ratio $R_{\sigma}$ of the cross sections measured in
the $\ell$+jets and dilepton channels and interpreted in terms of
charged Higgs decays of the  top quark, resulting in an upper limit on   ${\mathcal
  B}(t \rightarrow H^{+}b)$ for a leptophobic charged Higgs model.

With higher luminosity and ever improving techniques, the SM $t\bar{t}$ and
single top sector can be explored more and more precisely. 
Until the end of Tevatron's
Run~II about $6-8{\rm~fb}^{-1}$ of data are expected, yielding a lot of
opportunities to search for new physics.

\bibliographystyle{aipproc}   

\bibliography{sample}

\IfFileExists{\jobname.bbl}{}
 {\typeout{}
  \typeout{******************************************}
  \typeout{** Please run "bibtex \jobname" to optain}
  \typeout{** the bibliography and then re-run LaTeX}
  \typeout{** twice to fix the references!}
  \typeout{******************************************}
  \typeout{}
 }

\end{document}